\documentclass[10pt,conference]{IEEEtran}
\usepackage{color}
\definecolor{darkblue}{rgb}{0.0,0.0,0.6}
\definecolor{darkgreen}{rgb}{0,0.5,0}
\usepackage{listings}
\ifCLASSINFOpdf
  \usepackage[pdftex]{graphicx}
\else
  \usepackage[dvips]{graphicx}
\fi
\ifCLASSOPTIONcompsoc
\usepackage [caption = true, font = normalsize, labelfont=sf,textfont=sf]{subfig}
\else
\usepackage [caption = true, font = footnotesize] {subfig}
\fi

\IEEEoverridecommandlockouts\IEEEpubid{\makebox[\columnwidth]{978-1-5090-0223-8/16/\$31.00~\copyright~2016 IEEE \hfill} \hspace{\columnsep}\makebox[\columnwidth]{ }}

\begin{document}

%
\title{SDxVPN:   A Software-Defined Solution for  VPN Service  Providers}


\author{\IEEEauthorblockN{Behzad Mirkhanzadeh \IEEEauthorrefmark{1}, Naeim Taheri \IEEEauthorrefmark{2}, Siavash Khorsandi \IEEEauthorrefmark{3}}
\IEEEauthorblockA{\IEEEauthorrefmark{1}Faculty of Computer and Information Technology Engineering, Qazvin Branch, Islamic Azad University, Qazvin, Iran,}\IEEEauthorblockA{
 \IEEEauthorrefmark{2}Sharif University of Technology, \IEEEauthorrefmark{3}Amir Kabir University of Technology } \IEEEauthorrefmark{1}b.mirkhanzadeh@gmail.com,
\IEEEauthorrefmark{2}ntaheri@ce.sharif.edu, \IEEEauthorrefmark{3}khorsand@aut.ac.ir}


%


\maketitle

\begin{abstract}
BGP/MPLS IP VPN and VPLS services are considered to be widely used in IP/MPLS networks for 
connecting customers' remote sites. However, service providers struggle with many challenges to provide these services.   Management complexity, equipment costs, and last but not least, scalability issues emerging as the customers increase in number, are just some of these problems. Software-defined networking (SDN) is an emerging paradigm that can solve aforementioned issues using a logically  centralized controller for network devices.
In this paper, we propose a SDN-based solution called SDxVPN which considerably 
lowers the complexity of  VPN service definition and management. Our method eliminates complex and costly device interactions that used to  be done through several control plane protocols and enables customers to determine 
their service specifications, define  restriction policies and even ‌interconnect with other
customers automatically without operator's intervention. We describe
our prototype implementation of SDxVPN and its scalability evaluations under several representative
scenarios. The results indicate the effectiveness of the proposed solution for deployment to provide
large scale VPN services.\end{abstract}


%
\IEEEpeerreviewmaketitle

\section{Introduction}
Virtual private network (VPN) services are among the important services of carrier-grade service providers (SP). These services are provided for many customers and aim to connect customers' geographically distributed sites. Since IP/MPLS is dominant in the core of carrier class  networks, VPN services are realized using MPLS. While layer 3 VPNs between customers' sites are usually provided through ''BGP/MPLS IP VPN'' (Also called MPLS VPN) services, ''VPLS'' is the most famous service for providing layer 2 VPNs through the MPLS network. In this architecture, the service provider network is divided into two parts: the core region and the edge. On the edge, the provider network is connected to the customer's network via their provider edge (PE) devices. PE devices are connected to each other by the core network. In general, the core network consists of routers using MPLS technology in order to forward traffic among the PEs, and customer’'s sites are connected to the provider’'s network through customer edge (CE) devices \cite{14}.

We have been observing the workflow regarding provisioning and maintenance of  MPLS VPN and VPLS services in Iran's Telecommunication Infrastructure Company\footnote{http://www.tic.ir} for over a year and discovered major issues, also mentioned in some other related works. These issues are as follows:
\begin{enumerate}

\item[(1)]
\textbf {Management Complexity:}
MPLS VPN and VPLS services impose some serious management difficulties on service providers.  The distributed architecture of control plane causes complexity in configuration of the network done in a device by device manner. As a result, SP operators must learn the complicated and time-consuming VPN configuration process in order to provision and maintain  VPN services for many customers \cite{28}\cite{7}\cite{12}. These problems are intensified in a multi-vendor environment. Currently, some network management systems and configuration automation solutions such as \cite{4}\cite{5}\cite{6}\cite{7} and even some OSS/BSS solutions have been engaged in order to facilitate the management of these services. However, they are still tied to vendor specific commands and suffer from lots of complexity in their underlying configuration engines components. 
\item[(2)]
\textbf {Expensive devices:}
Since in the current architecture the control and data planes are vertically integrated, a considerable number of control functions must be implemented in PE devices (e.g. MP-BGP, LDP, IS-IS). In addition, customers'’ numerous IP prefixes (in MPLS VPN) and MAC addresses (in VPLS) must be maintained by PE devices. As a consequence, the network demands expensive and high-performance routers (e.g. Cisco 7600 series). Using many of these devices is not a cost effective solution for service providers regarding the growing number of customers and their excessive need for PE devices. 
\item[(3)]
\textbf {Scalability:}
Currently providers of MPLS VPN and VPLS services face some serious scalability concerns; for instance, PE devices provide a limited amount of memory for storage of MAC addresses and IP prefixes of numerous customers. Besides, running control functions (such as MP-BGP for MPLS VPN and maintaining full mesh of pseudowires among PEs for VPLS) puts a heavy load on PE devices, especially  when the number of PE devices increase according to growth of services. Although there has been some attempts to solve the scalability problems of MPLS VPN and VPLS services so far \cite{9}\cite{3}\cite{8}, they are just revamping existing architecture and not solving the  fundamental issue. 
\end{enumerate}

We believe that  software-defined networking (SDN) can solve these issues. SDN is an emerging network architecture that promises better network management by decoupling control and data planes. According to this architecture, the  data plane  turns into  a simple forwarder device consisting of flow tables to be filled by a logically centralized controller   which runs   network applications on top of itself \cite{15}. Data and control planes communicate through standard protocols among which OpenFlow \cite{16} is the most widely adopted one.
 
In this paper, we introduce our  software-defined approach called ''SDxVPN' by which the SPs can provide MPLS VPN and VPLS services. We show that our approach can solve  the aforementioned triple issues. First, by using SDxVPN each customer has a dedicated network application and can simply provision and maintain VPN services by himself. In other words, the service provisioning/maintaining processes, which impose lots of problems on SP's operators, can be fulfilled by the customer in a much easier way.  Second, by separating control and data planes, PE devices become simple forwarders which do not cost as much as current devices do. As a result, while SPs can use their existing devices, they are not obligated to buy expensive PE devices for scaling their networks. Third, we show that our solution is  scalable enough for handling numerous customers' services and it is not suffer form scalability issues which has been emerged due to the running of distributed control plane protocols on PEs. 

Figure \ref {fig1} illustrates the design perspective of  SDxVPN with a hypothetical carrier class network providing VPN services. Two different customers exist (distinguished by colors) each having different sites in different locations. The CE devices of these sites are connected to PE devices so as to have a VPN network realized by VPLS or MPLS VPN services. PE routers are changed to software-defined switches controlled by our logically centralized SDxVPN controller via OpenFlow. As a result,
PE to PE control plane protocols like MP-BGP and their required attributes such as Route Distinguisher (RD) and Route Target (RT) will be eliminated.
Moreover, each customer (\textit{C1} and \textit{C2} in the Figure) has a dedicated software-defined application, which operates on top of the SDxVPN runtime, for provisioning and managing his services. To achieve a practical solution, there are important challenges that need to be addressed:

\begin{figure}
\centering
\includegraphics [width = 2.5 in]{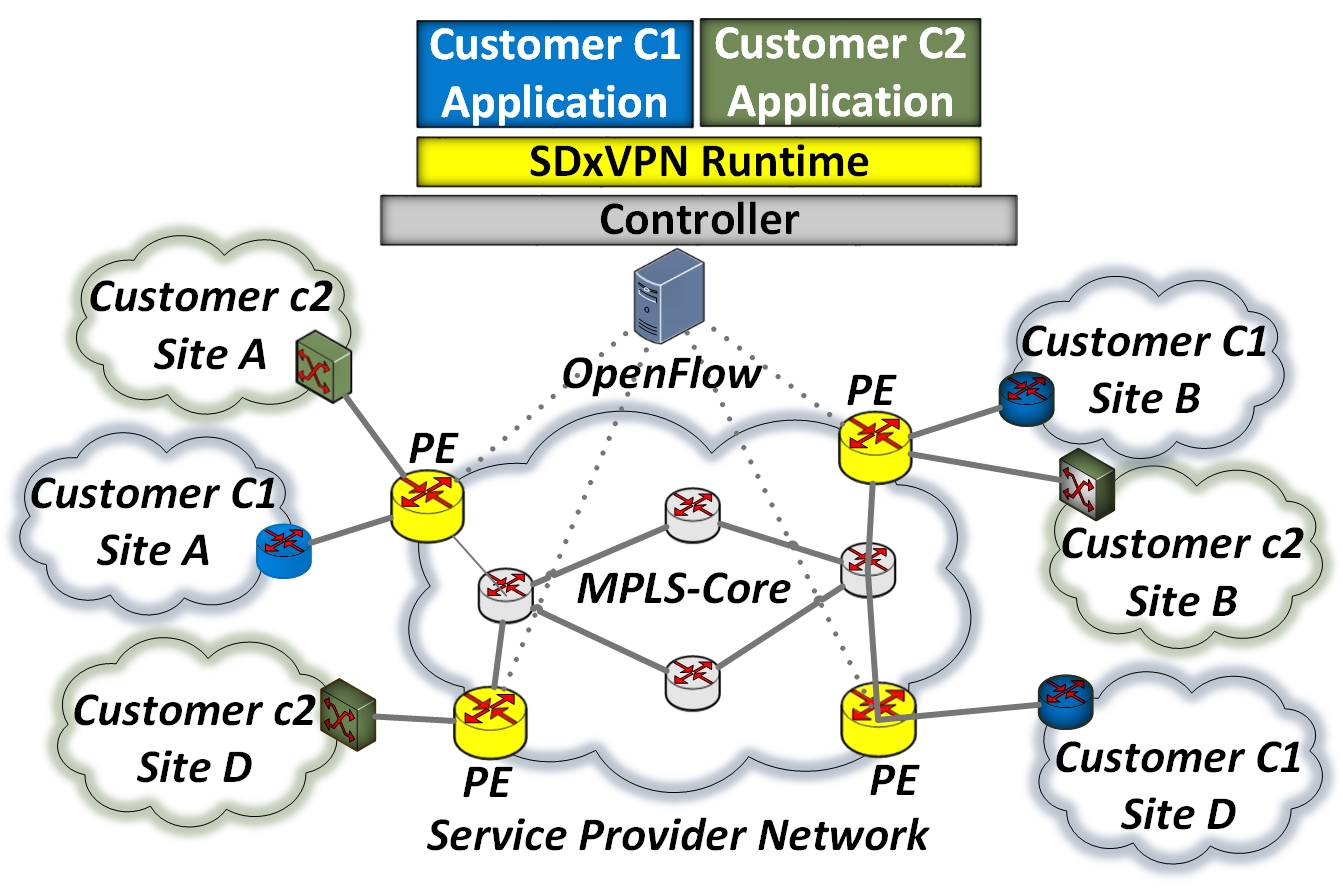}
\caption{Design perspective of SDxVPN.}
\label {fig1}
\end{figure}

\begin{itemize}
\item \textbf {Network Virtualization and Abstraction:} By providing each customer with an abstract network slice, they can have control over their services. We propose a method to slice SP network for every customer, thereby equipping them with dedicated applications which are able to provide new features (such as restrictive policies) that legacy MPLS VPN and VPLS offered with difficulty. These features can be configured by the customer himself without the  interference of service provider's operators. (Section \ref{sec:2})

\item \textbf {Hybrid Networking:}
Applying SDN to  a SP network is  preferably done in a gradual manner which results in a hybrid network consisting of SDN and non-SDN regions.  In 
Section \ref{sec:3}, we show how SDxVPN forms a hybrid network that enables software-defined PEs to interoperate with the MPLS core network.

\item \textbf {Scalability:}
Scalability has always been an important concern for carrier-grade service providers due to their large number of customers. Using SDN itself also has some scalability issues that should be considered. In Section \ref{sec:4}, we elaborate on our architecture and clarify how our design copes with such issues. \end{itemize} 

In the remaining sections, we describe a prototype of SDxVPN and evaluate its scalability in terms of flow table size (Section \ref{sec:6}). Finally, we present related works in Section \ref{sec:7} and conclude our work in Section \ref{sec:8}.

\section{Network Virtualization and abstraction}\label{sec:2}

\begin {figure*}[!t]
\centerline {\subfloat [ A Virtual Router realizes MPLS VPN.] {\includegraphics [width= 2.6 in, height= 1.65 in] {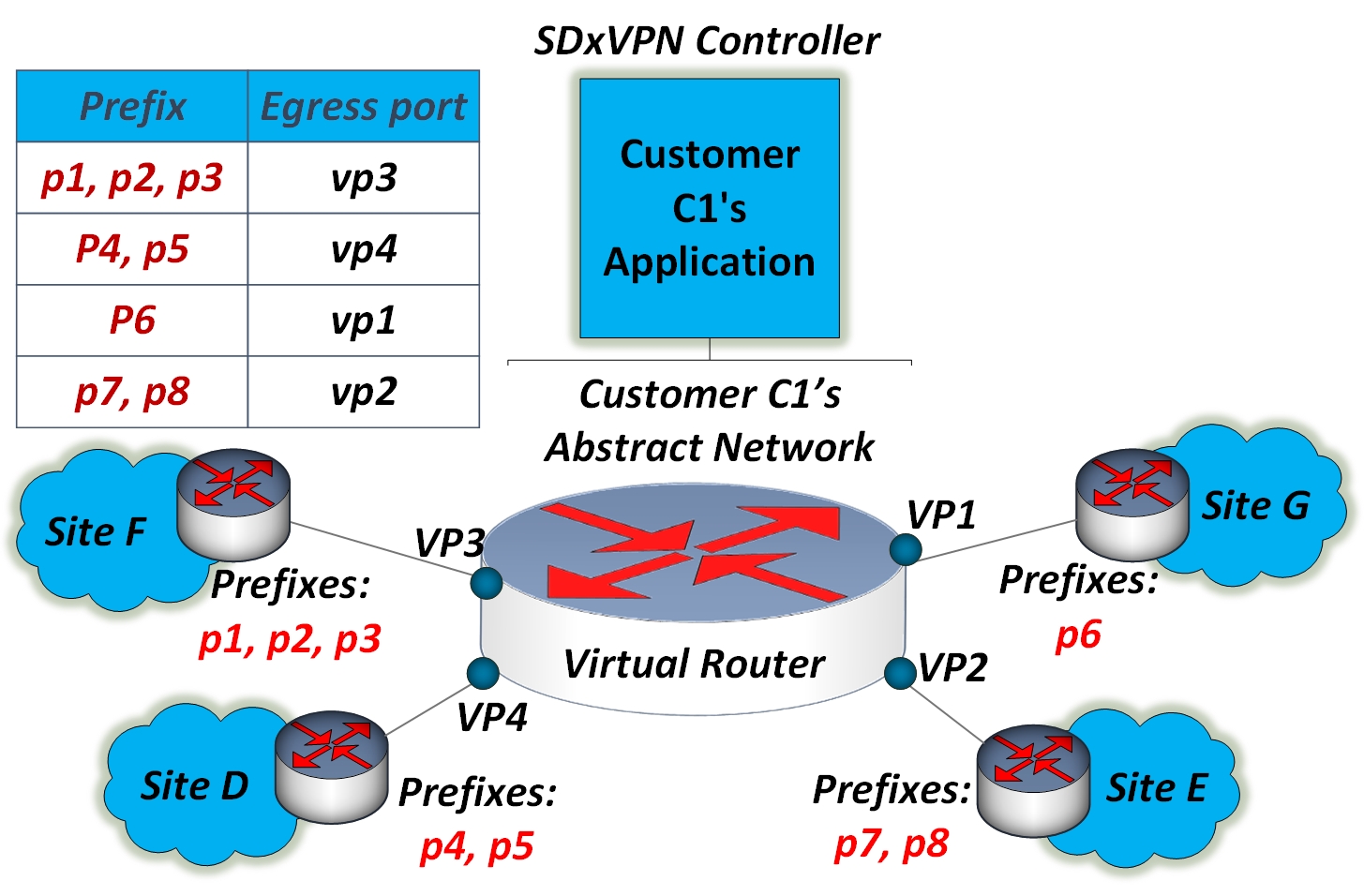}
\label{fig2-subfig1}}
\hfil
\subfloat [A Virtual Switch realizes VPLS. ] {\includegraphics [width= 2.6 in, height= 1.65 in]{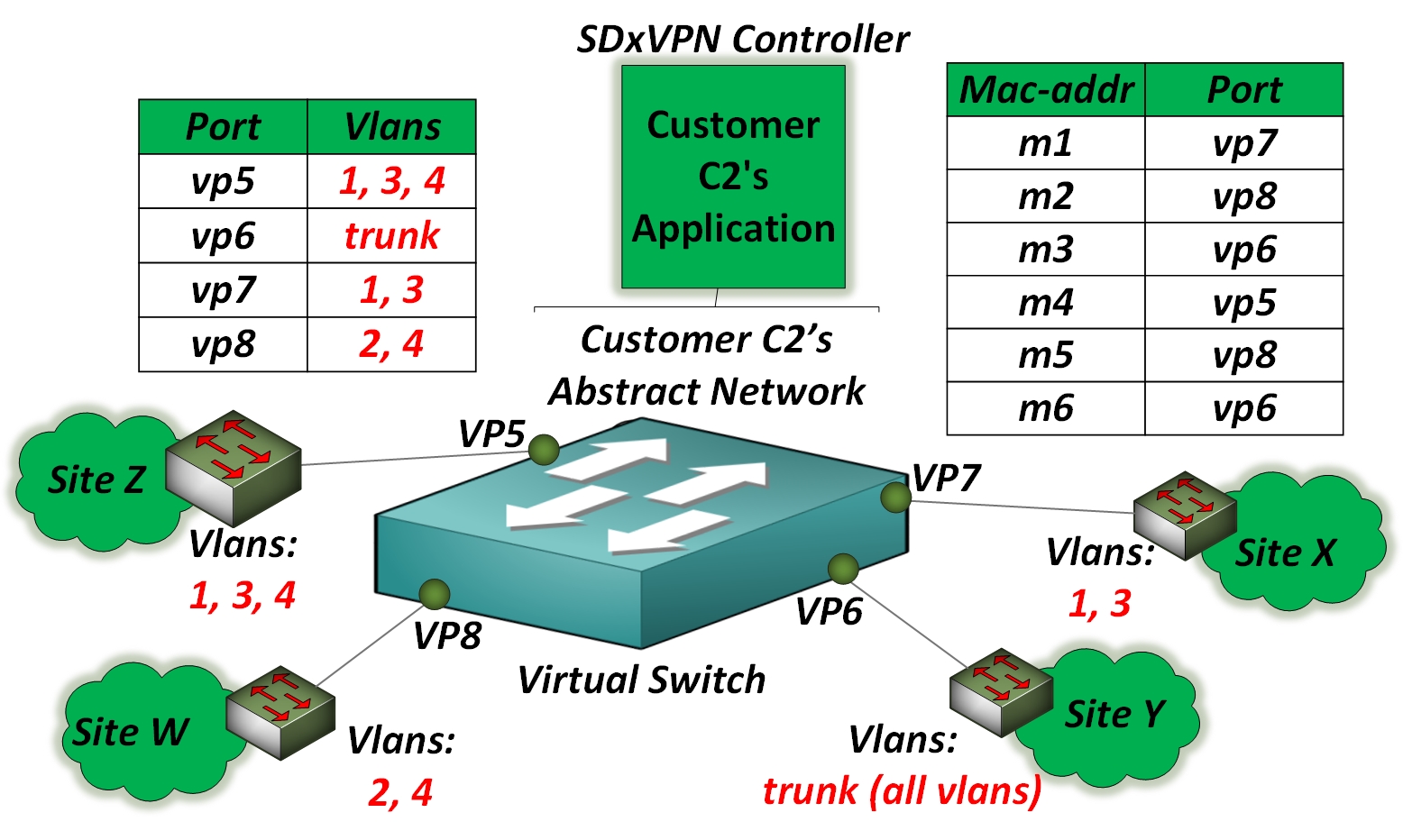}
\label{fig2-subfig2}}}

\caption{  Each service operates on its own virtual slice of the service provider network.}
\label{fig2}    
\end {figure*}

Customers prefer to have a flexible control over the VPN services they are purchasing. Independent management can be realized if customers are given their own virtual slice of the network to configure. Slicing the network in the current architecture is difficult due to the fact that data and control planes are integrated. Employing software-defined networking can facilitate network virtualization since virtualizing simple SDN switches can be done easier \cite{17}.

To enable customers to configure and modify their virtual networks, we have  designed a network virtualization method on the SDxVPN control plane. As illustrated in Figure \ref {fig1},  a separate application for each customer is installed on top of the SDxVPN runtime. These applications work on an abstract network which operates on a slice of the provider's network. For a customer who demands MPLS VPN service, SDxVPN runtime offers a virtual router. On the other hand, SDxVPN proposes a virtual switch for VPLS services. Each port on these virtual devices represents a real port of a PE device connected to the customer's CE device. Indeed, a virtual device with its ports give the customer an abstract view of the entire SP network.
The customer application has several features that will be elaborated in the following subsections.
\subsection{Service Specification}\label{sec2a}
SDxVPN enables customers to define VPN services in detail. Customers are able to define the services they wish in terms of a virtual router/switch operating on the ports they have ordered. Figure \ref {fig2} demonstrates network slices for two hypothetical customers, \textit{C1} and \textit{C2}.

\subsubsection {\textbf{ MPLS VPN}}

In Figure \ref {fig2}(a), a virtual router with basic routing functions and four virtual ports is presented. For determining MPLS VPN service specifications, customer \textit{c1}  should first assign an IP-address to each virtual port based on the IP-address range of the CE interface connected to it. Secondly, the customer should determine  the routing mechanism.
For example, in a simple scenario like this,  \textit{c1} can adopt static routing which requires
specifying prefixes manually for each site. Based on this specification, a routing table will be constructed. This routing table will then be translated into OpenFlow rules by SDxVPN runtime and will be installed on the PE devices mapped to the virtual ports. Following the example, packets destined for prefixes \textit{p1}, \textit{p2} and \textit{p3} must be forwarded to \textit{VP3}. Consequently, SDxVPN runtime configures the respected PE devices by the proper rules to perform the expected forwarding action.
In MPLS VPN services, since customers might prefer running their own desired routing protocol to advertise prefixes from one site to another, an implementation of common routing protocols on SDxVPN is necessary. To achieve this, we have employed Quagga  \cite{19} as our route engine in the virtual router. Using this engine, virtual routers will be enabled to co-operate with CE routers via various routing protocols like OSPF and BGP. We have designed an OpenFlow adapter for virtual routers that enables Quagga to exchange routing control messages with CE devices via OpenFlow enabled PEs  (explained in Section \ref{sec:4}).
 
\subsubsection {\textbf{VPLS}}

In Figure \ref {fig2}(b), the virtual switch manifests the realization of a VPLS service. Customer \textit{c2} can set each virtual switch port to operate on specific VLANs. For example, \textit{VP5} is adjusted to operate on VLANs \#1, \#2 and \#3 since site \textit{z} operates on these VLANs. On the other hand, \textit{VP6} operates on all VLANs (trunk) because the customer needs his connected sites to receive traffic from all VLANs. The table  presented on the left side of Figure \ref {fig2}(b) shows ports bound to VLANs. After this binding,  the  virtual switch must maintain a forwarding table (like a real switch) in order to forward  traffic  to the appropriate sites. For each virtual switch, we employed a learning mechanism to remember source MAC address and its associated ingress virtual port. Finally, like MPLS VPN, forwarding decisions will be dynamically translated to OpenFlow rules and will be set on the data plane by SDxVPN runtime. To clarify, after each site in a VPLS domain sends traffic to a PE device, if no rule corresponds to the sent frame, it will broadcast across all other customer’'s sites operating in the frame VLAN. Meanwhile, a copy of the frame will be sent to the controller for MAC-learning. Subsequently, in order to mitigate broadcasting the virtual switch installs proper rules on the PEs to determine the exact path to reach the learnt MAC address.

On the right side of Figure \ref {fig2}(b) a forwarding table for the virtual switch is illustrated. For instance, since the MAC address \textit{m2} has already been learned by the virtual switch, the frame destined for \textit{m2} will be forwarded to site \textit{w}. As another example, consider a broadcast frame (like an ARP request) or a frame destined for a MAC address unknown to the forwarding table, sent from a host within site \textit{z} having VLAN tag \#1. The frame will broadcast across all sites except \textit{w} because \textit{vp8} is not bound to VLAN \#1. Additionally, according to the learning mechanism, the source MAC address of the frame will be added to the forwarding table of the virtual switch. 

\subsection{Policy Specification}

In SDxVPN, customers can enhance their applications with policies to restrict VPN domains communications. Restrictions that would have been applied through configuration of several CE devices, now can be centrally defined within customer application. We have provided a simple xml-based language  by virtue of which customers can define policies to limit traffic entering the virtual switch/router. The structure of this language is as follows:

\begin{lstlisting}
<policy>
	<match name="{headerFiled-name}" 
		value="{field-value}"/>
	<apply>
		<virtualport name="{vPort-name}"  
			direction="{in/out}"/>
	</apply>
</policy>
\end{lstlisting}

In this language, each policy can be  applied to the traffic flowing in or out of the specified virtual ports and matched against defined packet header fields. Regarding Figure \ref{fig2}(a), customer \textit{c1} can restrict access to  a server with IP address 192.168.10.1 inside \textit{site F} from all sites except \textit{site D} :
\begin{lstlisting}
<policy>
	<match name="destinationIP" 
		value="192.168.10.1" />
	<apply>
		<virtualport name="vp1" direction="in"/>
		<virtualport name="vp2" direction="in"/>
	</apply>
</policy>
\end{lstlisting}

Policies of all customer applications will be accumulated and translated into OpenFlow rules with top priority and will be installed on the PEs' flow tables. 

\subsection{Multi-customer VPN}

Customers sometimes use a secure tunnel between one another's sites over Internet 
for special applications such as inter-banking transactions; However, already established VPN services of multiple customers could be extended for connecting their sites together in a more secure and economical 
fashion. One of the interesting features of SDxVPN is \textit{''Multi-Customer VPN''} by which multiple customers 
can create a private network and share their sites among each other without the SP's operator intervention. 
In other words, a Multi-customer VPN involves VPN domains that belong to different customers.
Introducing such compelling feature to the current MPLS VPN and VPLS services is very difficult.
For example,
sharing VPN domains among multiple customers in MPLS VPN services
makes the designing of \textit{Import \& Export RT} values quite challenging for SP's operators.

In SDxVPN we realize Multi-customer VPN by introducing another abstraction over the SP's network for both MPLS VPN and VPLS services.  
“The Shared Virtual Switch (SVS) and Shared Virtual Router (SVR) 
are provided for customers who want to form a private network altogether through VPLS and 
MPLS-VPN services respectively. The functionality of SVS and SVR 
is the same as the functionality of the virtual switch and router. 
The only difference is that the ports of these virtual devices belong to multiple customers 
and each of these customers have control over only the ports they own. 
We note that policies defined in the previous section can also be 
applied to SVS and SVR ports for defining further restrictions.

\section{hybrid networking}\label{sec:3}

Changing the entire SP’ network into a software-defined architecture may cause some problems. For example human resistance may happen because of the sudden changes in role and power \cite{15}. Accordingly, adoption of SDN should be performed in an evolutionary manner emerging as a hybrid network consisting of a SDN edge co-operating with the non-SDN core network (See Figure \ref {fig1}). Because of this matter, Label Distribution Protocol (LDP) \cite{18} must be implemented on SDxVPN to enable PE devices to send appropriate labeled traffic to the core network. Each PE device has its own instance of \textbf{"Core-Mediator"} component in the SDxVPN control plane consisting of LDP and Quagga route engine subcomponents. 

A Quagga route engine is also required in this step. Since LDP depends on the routing information base (RIB) provided by an IGP protocol (usually OSPF or IS-IS) that operates among the core network and PE devices. Quagga sets an IP address  for each interface connected to a core device and a router-id for the PE device  itself, then establishes adjacency with the core routers to exchange prefixes. The best route to all PEs are calculated and the next hop (a core router) is stored in the forwarding information base (FIB).

The SDxVPN LDP engine operates similar to a legacy router's LDP engine. It uses router-id determined by Quagga to communicate with adjacent routers. Initially, a LDP session is established with the adjacent routers, and a local label is assigned to the router-id. The locally generated label will be sent via label mapping massages to enable core routers to send traffic to the PE with the expected label. Simultaneously, label mapping massages will be received from adjacent core routers containing expected labels to forward traffic to other PEs. The received labels gradually fill out the local information base (LIB). In order to  reach a target PE, the LIB identifies a specific outgoing label for every distinguished next hop.
Applying the FIB on the LIB results in a Label Forwarding Information Base (LFIB) table in which each entry specifies the next hop and its respected label which determines the Label Switch Path (LSP) to reach a PE destination.

\section{Architecture design}\label{sec:4}

Applying SDN to a service provider network, as the size of the network increases, raises three important scalability issues:
\begin{enumerate}
\item[(1)] The control plane is potentially a bottleneck
if the data plane can not make forwarding decisions and imposes a heavy load on the controller by sending major part of traffic to it. \cite{15}\cite{20}.

\item[(2)] The process of flow setup requires interaction between the data plane (which consists of geographically distributed PEs) and the controller. This results in forwarding latency and degradation of QoS \cite{15}\cite{20}.

\item[(3)] Data plane resource limitation due to the huge number of flow rules installed on devices \cite{20}.
 
\end{enumerate}

We have considered these scalability issues in the SDxVPN architecture design. In order to tackle the first and the second challenges, we adopted a mechanism to install flow rules  proactively. To put it differently, instead of sending the first packet of each flow to the controller to make forwarding decision, the controller supplies the data plane with proper rules to make forwarding decisions earlier. Once the customer application determines service specifications, appropriate rules will be installed on PE devices to make forwarding decisions nearly independent of the controller. This method not only reduces the load on the controller, but also minimizes the effect of flow setup latency on QoS. To handle the third challenge, we employed a data plane efficiency strategy by means of which  flow tables  are divided into multiple stages and provide a dedicated table for each services. Using this mechanism, we made the rule table size of each service manageable and even reduced the number of flow entries in some cases.

We now describe the overall architecture of  both control and data planes in SDxVPN.
\subsection{Control plane}\label{sec:5}

\begin{figure}[!t]
\centering
\includegraphics [width = 2.9in]{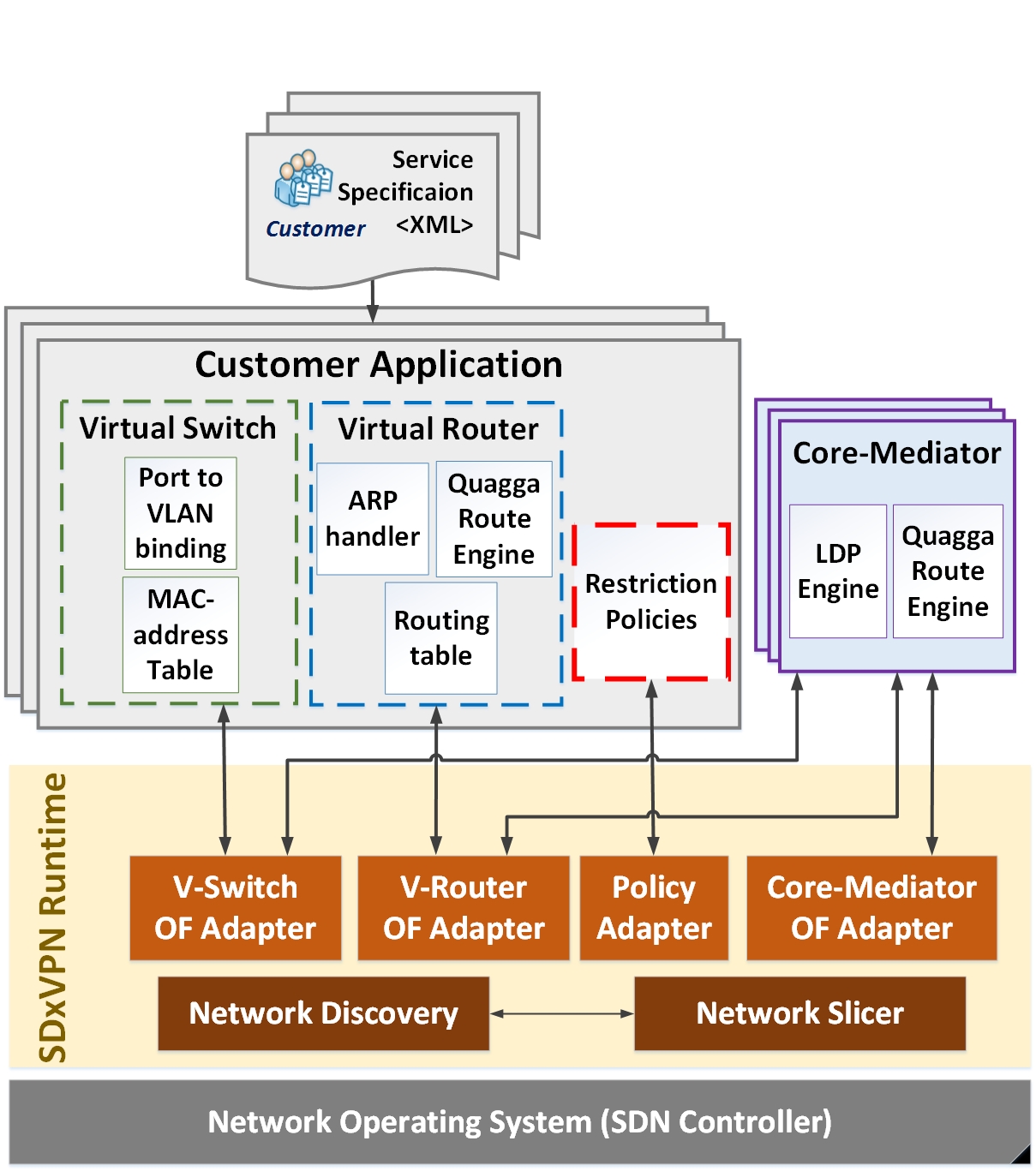}
\caption{SDxVPN Controller Architecture.}
\label{fig3}
\end{figure}

\begin {figure*}[!t]
\centerline {\includegraphics [width= 5.6 in] {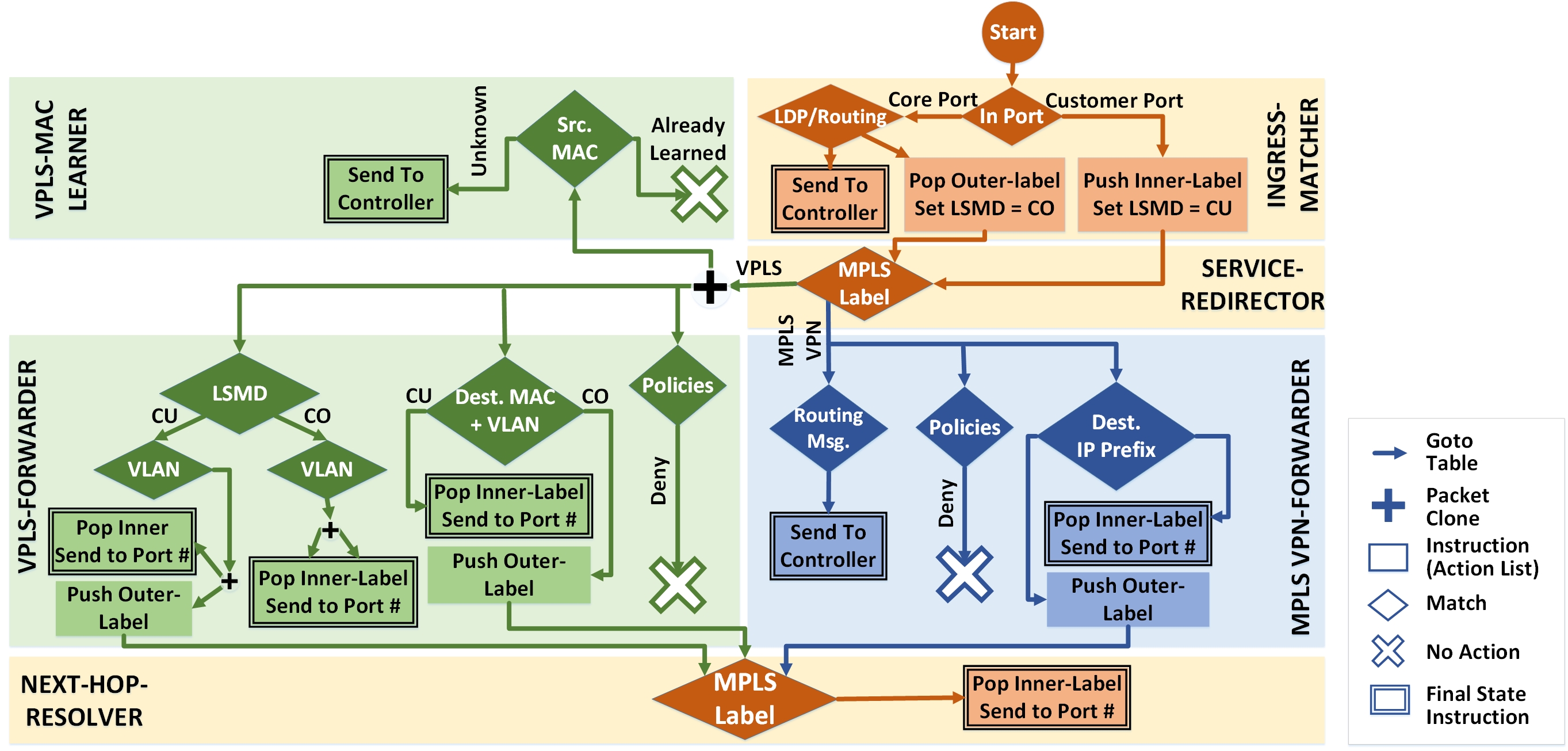}}
\caption{The Flowchart of SDxVPN Data Plane.}
\label{fig4}
\end {figure*}

In the control plane, we adopted a combination of component-based and layered architecture, a common approach in software-defined networking to manage complexity \cite{15}. SDxVPN architecture is represented in the Figure \ref{fig3} where Customer and Core-Mediator applications are implemented independent of the data plane protocol, running in parallel on top of the SDxVPN Runtime, and communicate with the lower layer using Restful APIs. SDxVPN runtime has three main responsibilities: slicing the network for higher-level applications, discovering the topology and providing each application with proper adapters for data plane communication.

In a typical scenario, each customer passes the service specification to his dedicated application in which
the appropriate forwarding tables of the virtual devices will be constructed.
Afterwards, the forwarding tables will be  passed to the OpenFlow adapters which are in charge of handling inter-communication between network applications and the OpenFlow PE switches.
Therefore, using our method, customers
will have direct control over their services and the intervention of SP's operators will be minimized.

The short description of the main components of SDxVPN architecture are as follows:

\textbf {Network Slicer \& Discovery.} The slicer component cooperates with "Network Discovery" module which is in charge of discovering software-defined PEs. The discovery can be done dynamically (through a protocol like LLDP) or manually (described through XML specification). After gaining topology of the network, the slicer component can determine the ports of PEs connected to core routers to exchange the control traffic (routing and LDP related traffic) between core routers and core-mediator applications.
The slicer also has a database of PE ports with unique hash-keys assigned to them. Customers who have purchased a port are given the hash-key to use it in virtual devices within their application.

\textbf {Customer Application.}
Customers define the specification of their services in the form of an XML-based language embedded within their applications. 
This simple language enables customers to describe the features and attributes of the \textit{Virtual Routers}, \textit{Virtual Switches} and \textit{Restriction Policies} (explained in Section \ref{sec:2}) according to  their service specification. 

\textbf {Core-Mediator.}
As described in Section \ref{sec:3}, the Core-Mediator component is implemented to realize the communication between SDxVPN and the MPLS core routers. It consists of LDP and Quagga route engine subcomponents for generating proper labels and constructing LSPs among PEs.

\textbf {OpenFlow Adapters.} The OpenFlow adapter is designed for several reasons: first, to provide  the proper abstraction and transparency of data plane complexity for applications defined on top of SDxVPN runtime. Second, to satisfy the need for a thin coordinator to handle communication between OpenFlow switches and applications.
In the current state we have considered four OpenFlow adapters:
\begin{enumerate}
\item[(1)] \textbf {Core-Mediator Adapter:} The core-mediator adapter is responsible for relaying LDP and routing related traffic from the "Core-Mediator" application to core routers and vice versa.\item[(2)] \textbf {Virtual Router Adapter:} The virtual router adaptor
has three main functionalities: 1) Exchanging ARP request/response between CEs and ARP handler subcomponent within the virtual router. 2) Passing the routing related traffic between the "Quagga Route Engine" and the CEs. 3) Updating flow rules based on the FIB formed in the "Routing-Table" subcomponent.
\item[(3)] \textbf {Virtual Switch Adapter:} It has two responsibilities: 1) Sending frames with unknown source MAC address to the "MAC Address Table" subcomponent for learning purposes. 2) Updating flow rules based on the "MAC Address Table" and the "Port To VLAN Binding".

\item[(4)] \textbf {Policy Adapter:} This adapter must convert policies that customers have defined in their applications to OpenFlow rules with the highest priority so that they can be applied to the traffic earlier than default rules.\end{enumerate}

\subsection{Data plane}

The SDxVPN data plane has two key properties: First, a separate table is allocated for each customer. Isolating customer specific forwarding rules makes them more manageable. It also avoids problems that might occur due to the limitation of flow table size by specifying a fixed table size for each service. Second, the adopted MPLS labelling strategy is quite similar to the traditional approach. Customer’s' traffic needs two labels in order to be delivered properly through the carrier network: the inner and the outer labels. The inner label also known as the "service delimiter tag" distinguishes the traffic of each service from the others to avoid address overlapping. The outer label, assigned by LDP module, is used to forward traffic through the MPLS core to the desired egress PE.

Figure \ref{fig4} demonstrates the whole process which goes through the data plane of SDxVPN in a flowchart. Dim big rectangles in the background represent flow tables which are distinguished in three different colors: The yellow tables, namely Ingress-Matcher, Service-Redirector and Next-Hop-Resolver, handle general routines used for all services; VPLS-Forwarder and VPLS-MAC-Learner are colored in green and are VPLS specific tables; and finally, the sole blue table is the MPLS-VPN-Forwarder and is dedicated to implement routing for the MPLS VPN service.  Decision symbols (diamonds) represent a match and the small rectangles represent an instruction (an action list), when the rectangle has a double border-line this indicates the final state of the pipeline. An arrow line pointing to another table means a "goto table" instruction. In addition, we have added two new symbols to the chart: the cross symbol means no action (drop packet) and the plus symbol indicates that a separate copy of the packet flows through each exiting arrow.

In Ingress-Matcher, the incoming packet will be matched to an ingress port. If it is coming from customer-side ports, it will be labelled with the service delimiter tag and the least significant bit of metadata (LSMD) of the OpenFlow packet will be set to zero (CU) and forwarded to the Service-Redirector. But if the ingress port corresponds to core-side ports, there are two different possibilities: 1) the packet corresponds to LDP or routing messages, so it will be sent to the controller immediately. Or, 2) the packet comes from other sites, therefore its  outer-label will be popped and the LSMD flag will be set to one (CO) and passed to the Service-Redirector. 

The Service-Redirector has the responsibility to redirect traffic to the service specific table based on the service delimiter tag. For MPLS VPN, the traffic will be redirected to the MPLS-VPN-Forwarder table. For VPLS, in a more complicated scenario the packet will be duplicated and one copy will be sent to the VPLS-MAC-Learner and the other to the VPLS-Forwarder.

In MPLS-VPN-Forwarder, the rules are classified in three categories with different priorities. The category with the highest priority matches the routing messages coming from CEs to send them to the controller. The policy matching category comes next; it drops packets according to the restriction rules defined by the customer. The least priority belongs to forwarding logic, which matches packets against destination IP prefixes calculated by the virtual router of customer application. If the destination belongs to the sites connected directly to the switch, The inner label will be popped and it will be forwarded to the appropriate port. Otherwise, an outer-label  corresponding to the target PE will be assigned to the packet. Thereafter,  this packet will be sent to the Next-Hop-Resolver.

In the VPLS-Forwarder, there are categories with different priorities too.  The policy matching category applies and filters undesired traffic, first.  The next priority belongs to the MAC address aware forwarding, which matches packets against VLAN and destination MAC addresses and determines the destination. Obviously, for destination sites directly connected to the switch, we simply pop the inner-label and steer the packet towards the proper port. For external sites, on the other hand, it requires to push the outer-label and send it to the Next-Hop-Resolver. Finally, the least priority happens in the condition where either the traffic is a broadcast or fails to match with MAC address aware forwarding rules. In such cases, several copies of the frame must be created and sent to the appropriate sites operating in the frame VLAN. For both directly connected and external sites, same action as previously explained above will take place. Notice that by using the LSMD flag we set earlier, which indicates whether the traffic has come from core-side ports or not, we apply the split horizon principle to avoid loops \cite{14} (prevent sending traffic back to core network). 

To minimize broadcasting, virtual switches need a learning mechanism to install proper flow rules on the VPLS-Forwarder table.  A  simple  implementation of  the learning  process  would be sending the source MAC address and ingress port of the incoming frames to the SDN controller, but in our case we cannot afford  this  type  of  load.  In order to handle this, we have designed a VPLS-MAC-Learner table to make the learning process more efficient and reduce the traffic that causes heavy load on the controller. When the controller learns a MAC address, it registers the source MAC address to this table. Incoming frames will be matched to this table, so the datapath does not forward frames with already known MAC addresses to the controller. As a result, only unknown frames (the first frame in each traffic burst) will be sent to the controller and the load on the controller will be minimized.

Overall, we proactively set up rules on the PE devices to make forwarding decisions as independent as possible from the controller. The packets redirected to the control plane are limited to LDP/routing messages and unknown source MAC addresses. Furthermore, the pipelining strategy used in the data plane results in a lower number of rules \cite{15}. Besides, by virtue of separate service tables, more manageable flow tables are achieved; the number of flow rules for each service can be restricted separately to prevent Flow-table explosion in the case of PE devices with limited memory. Thus, we believe that scalability concerns (discussed at the beginning of this section) are properly addressed in SDxVPN.

\section{Implementation and  evaluation}\label{sec:6}

We have implemented a prototype of SDxVPN using Floodlight \cite{21} and tested it via Mininet \cite{22}. Floodlight is an enterprise-class Java-based OpenFlow controller that supports OpenFlow 1.3. It suits our needs for MPLS labelling and matching. In our implementation, we have worked under two major assumptions:
\begin{enumerate}
\item[(1)] We have eliminated the core part of the network and instead directly linked each pair of PEs, for two reasons: The first reason is that the current version of OpenVSwitch\footnote{OpenVSwitch 2.3.2 as of Aug 20th 2015} does not support pushing more than one MPLS label. This problem is also mentioned in \cite{14} where they have used a VLAN tag as the outer label. We cannot use this strategy either, since SDxVPN at the same time supports VPLS service in which customer VLANs must remain unchanged.
The second reason is that communicating with core devices through LDP and Quagga engine does not have much effect on our evaluation which is based on rule table size because these features do not add many rules to the PE devices.
\item[(2)]

In order to emulate VPN domains in the Mininet bed, instead of using real devices for CEs, we used simple Mininet hosts with several virtual network adapters. The primary adapter of the host is connected to the PE, and the secondary adapters represent hosts inside a site.
\end{enumerate}

As for testing and verification of SDxVPN, we defined several MPLS VPN and VPLS services that have overlapping addresses.
After OpenVSwitches connected to the controller, proper rules to realize virtual devices functionalities were successfully installed on the datapath assuring \textbf{OpenFlow Adapters} worked as expected.
We checked accessibility of customers' sites within each service, and also the isolation of both VPLS and MPLS VPN services to make sure the \textbf{Slicer component} partitioned the network for each service correctly.
For VPLS services, broadcast packets like ARP successfully transferred through PE devices.
To verify the functionality of the \textbf{VPLS-MAC-Learner} we generated random traffic between hosts at a fixed rate and observed \textit{packet\_count} of broadcast rules. We observed after a short while that the \textit{packet\_count} remained unchanged which illustrated the correctness of the VPLS-MAC-Learner.
In order to test multi-customer VPNs we put sites of different services together in the form of a multi-customer VPN. Finally we tested the policy feature by adding restrictive rules for packets (e.g Rejecting FTP traffic destined to a specific site).
 
\subsection{Evaluation}

\begin{table}[!t]
\renewcommand{\arraystretch}{1.3}
\caption{Different scales of a hypothetical service provider.}
\label{table1}
\centering
\begin{tabular}{|c||c||c||c|}
\hline
\bfseries Scale  & \bfseries \vtop{\hbox{\strut Number of PEs}} & \bfseries Number of Services & \bfseries \vtop{\hbox{\strut Average Number} \hbox{\strut  of sites per service }}\\
\hline
1 & 4 & 40 & 6\\
\hline
2 & 8 & 70 & 10\\
\hline
3 & 10 & 100 & 15\\
\hline
4 & 12 & 200 & 20\\
\hline
5 & 16 & 300 & 30\\
\hline
\end{tabular}
\end{table}

There are three major scalability issues which need to be addressed in our SDN solution: load management on the controller, latency, and table explosion.  Regarding the first and the second issues, as we explained earlier in section \ref{sec:4}, we have proactively installed proper rules on the PE devices to make forwarding decisions as independent as possible from the controller. The only packets redirected to the control plane are LDP/routing messages and unknown source MAC addresses which seldom occur.

With regard to the third issue, we conducted an experiment and presented a numerical evaluation of the flow table size in various conditions. Table 1 shows hypothetical service providers in five different scales. For example, the first row which is the smallest scale represents a service provider with 4 PEs; there are 40 VPN services running each connecting 6 sites on average. In our experiment we assumed that the number of VPLS and MPLS VPN services are equal. There are conditions for each site like number of prefixes for MPLS VPN and number of MAC addresses within each site for VPLS that might affect the number of rules as well. Based on these conditions, we investigated three different scenarios: 1) Each site contains 4 IP prefixes or 30 MAC addresses. 2) Each site contains 6 IP prefixes or 40 MAC addresses. 3) Each site contains 8 IP prefixes or 60 MAC Addresses.

In addition to the previous assumptions, we hypothesized that each customer has specified 5 restriction policies for each site. Also, we presumed within each VPLS service there are 30 different VLANs. In addition, due to the bursty nature of traffic in the network we generated traffic between CEs such that hosts fluctuated between active and passive states. Consequently, "VPLS-Forwarder" and "VPLS-MAC-Learner" tables kept approximately half of the MAC addresses corresponding to each service in every time frame (Because of the soft timeout of the flow rules).

\begin{figure}[!t]
\centering
\includegraphics [width = 3.4in]{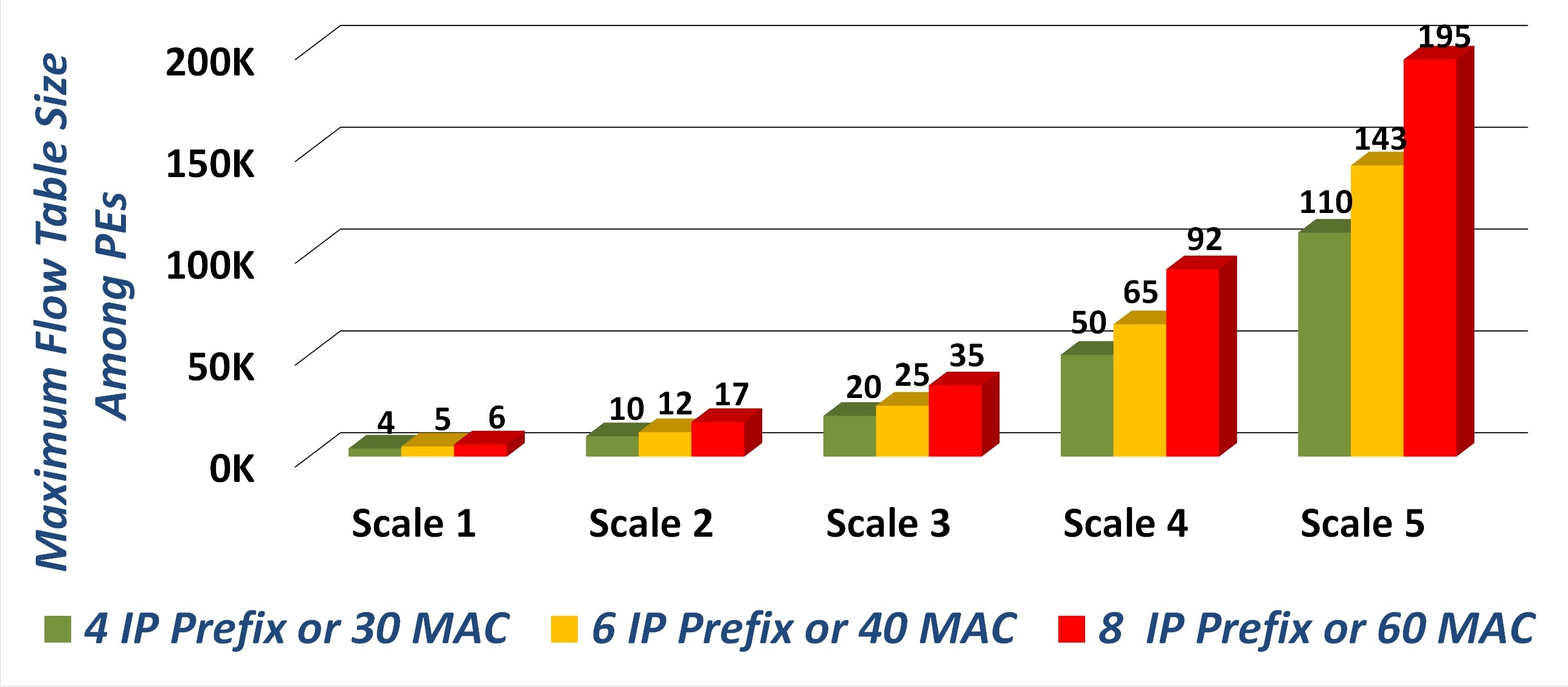}
\caption{Maximum flow table size among PEs for a hypothetical service provider in different scales and in various scenarios.}
\label{fig5}
\end{figure}

To benchmark our method in a difficult situation we connected a forth of the sites for each service to a single PE device to have a device with the largest number of rules. We ran our prototype for every scale under each mentioned scenario which resulted in 15 trials. Figure \ref{fig5} illustrates the flow table size with maximum entries among all PEs within each trial. In the worst case scenario, the bottleneck PE device reached 195K rules which is considerably less than the capacity of some currently manufactured OpenFlow devices like NoviFlow \cite{27} which supports up to 1 million flow entries. Note that the capacity of flow-tables of OpenFlow devices are increasing at a fast pace \cite{15}.
Furthermore,  in realistic deployments an extra management strategy can be applied to limit the number of flow rules to a fixed number for each service.

\section{Related Works}\label{sec:7}

In spite of  the eagerness for SDN, there have been few efforts to apply it in MPLS networks and associated VPN services. In \cite{10} the authors have mentioned some of  the issues regarding MPLS networks and briefly described their experience using SDN for MPLS traffic engineering. In \cite{11}  they added the   MPLS VPN   feature to their previous work. However, lack of details  and ignoring  layer 2 VPN services such as VPLS are some of the drawbacks of their work.
In addition, another group has tried to use SDN in MPLS networks and MPLS VPN services \cite{12}\cite{13}. In particular, they provide an XML-based language for specification of customer VPN services and then setup OpenFlow enabled PE and P devices with appropriate rules. Despite providing more details in comparison with the previously mentioned works, there are still  downsides to this work. The absence of network virtualization and abstraction in the control plane for facilitating network management, disregarding VPLS service and skipping the needs for routing protocols (for co-operating with CE devices)  are some of these weaknesses. 
 
There has been several efforts trying to gain a hybrid network consisting of
both SDN and non-SDN portions working together, like our solution.
In \cite{24} the authors have designed an IP/SDN architecture called Open Source Hybrid IP/SDN (OSHI) and proposed a node which can provide VPN services in a hybrid network. The major difference between this work and ours is that the main focus of OSHI is on the details of the IP/SDN data plane rather than the control plane and network management aspects.
In \cite{26} another OpenFlow enabled node is represented that uses Quagga and LDP protocol to perform MPLS labeling. However, their control plane is still integrated to the data plane and is not extracted from the device according to the standard SDN\ architecture.  There are two other significant works which are not directly related to MPLS networks and VPN services  but inspired us to propose a hybrid solution. Routeflow \cite{25}   proposes virtual routers (mirror of physical OF devices) on top of the controller in order to provide routing functions in hybrid networks.
A remarkable deployment of hybrid networking in large-scale networks has been done in Google B4 \cite{1}.
They migrated their backend network to SDN architecture and accordingly decreased their costs and simplified traffic engineering in their network. 

\section{Conclusion}\label{sec:8}
SDxVPN is a software-defined networking solution for service providers offering MPLS VPN and VPLS services. It eases management complexity, cuts expenses imposed by traditional non-SDN devices, and promises to solve the scalability issues that service providers have to deal with. In SDxVPN, customers are equipped with a dedicated network application through which they can express  service specification, define restriction rules and share their sites with other customers without SP operator's effort. SDxVPN is a hybrid network solution that enables OpenFlow  edge devices to co-operate with the non-SDN core portion, and allows service providers to migrate to a software-defined architecture gradually. We addressed scalability concerns in the design of both control and data planes and prepared  a prototype of our solution then evaluated it in terms of rule table size. The results indicated that SDxVPN is    scalable enough for even large-size service providers.

In future works, we intend to add some other services like Multicast VPN  and also redesign SDxVPN to work as a distributed controller to support service providers that expand over more than one country. In addition, we are interested to extend our solution to  core-routers and propose a Software-defined solution for the entire service provider network to simplify   MPLS traffic  engineering  in the core of the network.


\section*{Acknowledgment}

We would like to thank Telecommunication Infrastructure Company of Iran (TIC)
and IIS Center of Sharif University of Technology staff
that supported our research and let us observe the provisioning/maintaining work-flow of VPN services. 



%

\bibliographystyle{IEEEtranS}

\bibliography{SDVPN-cites}

\end{document}